\documentstyle[12pt]{article}

\def\pr{\partial}\def\la{\lambda}
\def\tb{\tilde{\beta}}

\title{Dispersionful analogues of Benney's equations and $N$-wave systems.}

\author{B. Enriquez, A.Yu. Orlov, V.N. Rubtsov}

\date{}

\begin{document}
\newtheorem{proposition}{Proposition}
\maketitle

\begin{abstract}
We recall Krichever's construction of additional flows to Benney's
hierarchy, attached to poles at finite distance of the Lax operator.
Then we construct a ``dispersionful'' analogue of this hierarchy, in
which the role of poles at finite distance is played by Miura fields. We
connect this hierarchy with $N$-wave systems, and prove several facts
about the latter (Lax representation, Chern-Simons-type Lagrangian,
connection with Liouville equation, $\tau$-functions).
\end{abstract}

\section{Introduction.}

It was realized recently that topological field
theories (TFT) are closely connected with integrable hierarchies. The
starting point of this connection were the Witten conjectures (proven by
Kontsevich) that the free energy of 2-dimensional (2D) gravity is the
logarithm of the KdV $\tau$-function, specified by the string equation.

Later, Dijkgraaf, E. and H. Verlinde, and Witten (\cite{DVVW})
showed that classes of 2D
TFT's are classified by solutions of a system of nonlinear partial
differential equations
bearing on some function (primary free energy of the theory). This
system is abbreviated WDVV; various classes of solutions to it were found by
Dubrovin and Krichever (\cite{Du}, \cite{KrENS}), using the integrable
systems arising from problems of isomonodromic deformations.
One of their results is that a class of WDVV solutions is
provided by $\tau$-functions for semiclassical limits of Hamiltonian
integrable systems.

$\tau$-functions of semiclassical (or dispersionless) limits of Lax-type
integrable hierarchies were studied in the papers of Krichever
(\cite{KrENS}) and Takasaki and Takebe (\cite{TT}), and generalized
$W$-constraints for these systems were computed. These results
inspired the sudy of dispersionless hierarchies and their relations with
other solvable systems, in particular of hydrodynamic type (\cite{DN}).
Among the long-known dispersionless systems of equations were the Benney
equations, introduced in \cite{Be}, which model long waves with a free
boundary. They were studied from the Hamiltonian point of view by
B. Kupershmidt and Y. Manin (\cite{KM}) and their relations with the
Kadomtsev-Petviashvili system were established in \cite{L}.

I. Krichever, in his talk at the 1994 Alushta conference, introduced a
dispersionless-type hierarchy, that may be viewed as a multiple poles
analogue of the Benney equations
(he also introduced an elliptic generalisation
of these sytems). The Lax operator of this system is a rational
function on ${\bf C}P^{1}$, with singularities at infinity and at some
points at finite distance; in addition to the usual KP flows,
corresponding to the pole at infinity, there are flows attached to the
points at finite distance. Krichever posed the problem to find
dispersionful analogues to these systems.

We solve this problem in the case where all poles at finite distance,
are simple. The resulting systems are expressed in Lax form; we call
them Krichever systems. The Lax
operator takes the form $AB^{-1}$, $A$ and $B$ differential operators.
We use the
idea that the evolution operators attached to poles at finite
distance are classical analogues of ``quantum'' Lax operators, written
using Miura fields of $B$.
We show that the resulting vector fields commute pairwise. We also study
the (bi-)Hamiltonian  properties of this system.
Our results can be considered as a generalization of results of
\cite{Di}, \cite{BX}, \cite{MR}, \cite{AFGZ}.

In the last part of the paper, we study the relation of the Krichever
systems with the $N$-wave systems (\cite{MZ}; they also appear
in \cite{Du}, in the context of WDVV equations). We show that the
solutions to the $N$-wave equations can be constructed from solutions to
the degree $1$ flows of the Krichever
hierarchy. We also show that
the $N$-wave system can be expressed as a commutation condition on
certain integro-differential operators; we give a Chern-Simons-like
Lagrangian for the $N$-wave system; we show a relation with a system
of ``connected'' Liouville equations (i.e. fulfilled for each pair of
indices); we constuct $\tau$-functions for the $N$-wave systems and
connect them with $\tau$-functions for KP systems. The question of
integration of the Krichever systems remains open; we make some
observations in section 4, indicating that it should be treated using
methods of \cite{Krconstraints}. We hope to return to this question
elsewhere.

\section{Extended Benney systems.}

  The phase space of these systems consists of the set of smooth maps
from $S^{1}= {\bf R}/{\bf Z}$ to the space of rational functions on
${\bf CP}^{1}$ of the form
$$
E(p) = p^{n} + \sum_{k=0}^{n-1}a_k p^k  +
\sum_{\alpha=1}^N \sum_{i=1}^{i_{\alpha}}\frac{a_{\alpha,i}}{(p-p_{\alpha})^i},
\quad	(n,i_{\alpha}\ge 0).\eqno (1)
$$

Viewing $E(p)$ as the symbol of a pseudodifferential operator, we can
define following \cite{LM}  a bracket operation on the space of maps
from $S^{1}$ to all rational functions on ${\bf CP}^{1}$, by degeneration
of the Lie bracket. We denote it by $\{,\}$. So $\{A(p,x),B(p,x)\} =
\frac{\partial A}{\partial p}\frac{\partial B}{\partial x} -
\frac{\partial A}{\partial x}\frac{\partial B}{\partial p}$.
Let us consider then the rational functions $E_{\infty,i}(p),
E_{\alpha,i}(p), i\ge 1$,
associated to $E(p)$, given by $E_{\infty,i}(p)=(E(p)^{i/n})_{+,\infty},
E_{\alpha,i}(p)=(E(p)^{i/i_{\alpha}})_{+,\alpha}$.
Here the powers of $E(p)$ are understood as formal expansions
near the points
$\infty$ and $p_{\alpha}$ respectively, and we set as usual
$$
(\sum_{s\le s_{0}}p^{s}u_{s})_{+,\infty} = \sum_{0\le
s\le s_{0}}p^{s}u_{s},
$$

and
$$
(\sum_{s\le s_{0}}\frac{u_{s}}{(p-p_{\alpha})^{s}})_{+,\alpha} =
\sum_{1\le s\le s_{0}}\frac{u_{s}}{(p-p_{\alpha})^{s}}.
$$

Define then flows $\partial_{\infty,i},\partial_{\alpha,i}$, by

$$\partial_{\nu,i}(E(p))=\{E_{\nu,i}(p),E(p)\}, \quad
\nu=1,...,N,\infty,\quad i\ge 0
\eqno (2)
$$

It can be shown that
this defines vector fields on the phase space. (For example,
the immobility of the pole at $\infty$ is connected to the fact that the
leading coefficient of $E(p)$ is $p^{n}$,
whereas the non-constantness of $x \mapsto a_{\alpha,i_{\alpha}}(x)$
induces the motion of $p_{\alpha}(x)$, i.e. its non constantness as a
function of time, $x$ being fixed.) Moreover, these vector fields commute.
Conserved quantities are in addition to the usual KP conserved quantities
$H_{\infty,i}$
$$
   H_{\infty,i}={\rm tr}_{\infty}(E(p))^{i/n}=
	 \int\limits_{S^{1}}{\rm res}_{\infty}(E(p))^{\frac{i}{n}+1}\,dx,
$$
$$
	 H_{\alpha,i}={\rm tr}_{p_{\alpha}}(E(p))^{i/i_{\alpha}}=
	 \int\limits_{S^{1}}{\rm res}_{p_{\alpha}}
(E(p))^{\frac{i}{i_{\alpha}}+1}
\,dx, \eqno (3)
$$

where in the last formula $E(p)$ should be expanded in formal series near
$p_{\alpha}$.

These systems were introduced by I. Krichever in his talk at the  1994
Alushta conference (\cite{KrAlushta}). They are a generalization of the Benney
systems (where all poles in expression (1) are simple),
introduced in \cite{Be}.

{\bf Remark.} It is possible to define, on the dispersionless phase
space, a compatible family of
Poisson structures attached to the poles at finite distance, in
the same fashion as the degenerations of the Adler-Gelfand-Dickey
structures are attached to the pole at infinity. But it seems difficult
to extend this construction in the dispersionful situation.

\section{Dispersionful analogues.}

We will construct analogues of the systems of section 2, in the case where
all $i_{\alpha}$'s are equal to $1$. In that case, we replace $p-p_{\alpha}$
by some operator $\partial + \phi_{\alpha}
=\exp(\varphi_{\alpha})\partial\exp
(-\varphi_{\alpha})$, and $a_{\alpha}(p-p_{\alpha})^{-1}$ by the operator
$a_{\alpha}\exp(\varphi_{\alpha})\partial^{-1}\exp(-\varphi_{\alpha})$;
so that the analogue of $E(p)$ should be some operator of the form
$$
{\cal L}= \partial^{n} + \sum_{k=0}^{n-1} p_{k}\partial^{k} +
\sum_{i=1}^{N} \varphi_{i}\partial^{-1}\psi_{i},\eqno (4)
$$
$p_{k},\varphi_{i},\psi_{i}$ are some functions. So the phase space will be
$$
{\cal P}=\{(P,\varphi_{i},\psi_{i})_{i=1,...,N}, P=\partial^{n} +
\sum_{k=0}^{n-1}p_{k}\partial^{k}, p_{k},\phi_{i},\psi_{i}\in C^{\infty}
({\bf R})\}.
$$
We have a mapping
$$
{\cal P}\rightarrow {\cal L}_{n}=\{\partial^{n}+ \sum_{k\le
n-1}u_{k}\partial^{k},
u_{k}\in C^{\infty}({\bf R})\}
$$
${\cal P}$ admits a $GL_{N}({\bf C})$-action by
$$
g\cdot (P,\varphi_{i},\psi_{i})=(P,(g^{-1})_{ji}\varphi_{j},g_{ij}\psi_{j}),
$$
and the fibres of the mapping ${\cal P}\to{\cal L}_{n}$
are generically the orbits of the action of $GL_{N}({\bf C})$.
Note also that
the part of ${\cal P}$ formed of the $(P,\varphi_{i},\psi_{i})$ s.t.
the Wronskian $W(\varphi_{i})$ (respectively, $W(\psi_{i})$) is invertible,
maps to the subset of ${\cal L}_{n}$ formed of the operators $AB^{-1}$
(respectively, $B^{\prime-1}A^{\prime}$), with
$A,A^{\prime}\in \{\partial^{n+N}+\sum_{0\le i<n+N}a_{i}\partial^{i}\},
B,B^{\prime}
\in \{\partial^{N}+\sum_{0\le i<N}b_{i}
\partial^{i}\}$; it is enough to take for
$B^{*}$ (respectively, for $B^{\prime}$) the operator with kernel
$\oplus_{i=1}^{N}{\bf C}\psi_{i}$,
(respectively, $\oplus_{i=1}^{N}{\bf C}\varphi_{i}$); the involution $B\mapsto
B^{*}$ is defined below.

We now show:

\begin{proposition}
$$
{\cal L}^{p}=({{\cal L}^{p}})_{+} + \sum_{i=1}^{N}\sum_{k=0}^{p}[{\cal L}^{k}
\varphi_{i}]\partial^{-1}[{\cal L}^{*(p-k-1)}\psi_{i}],\eqno (5)
$$
or equivalently
$$
(1-\lambda{\cal L})^{-1}=(1-\lambda{\cal L})_{+}^{-1}+\sum_{i=1}^{N}
[(1-\lambda{\cal
L})^{-1}\varphi_{i}]\lambda\partial^{-1}[(1-\lambda{\cal
L}^{*})^{-1}\psi_{i}].\eqno (6)
$$
Here ${\cal L}^{*}$ denotes the involution
$(\partial^{m}+\sum_{i<m}a_{i}\partial^{i})^{*} =
(-\partial)^{m}+\sum_{i<m}(-\partial)^{i}a_{i}$, we set
$(\sum_{i\le m}x_{i}\partial^{i})_{+}=\sum_{0\le i\le
m}x_{i}\partial^{i}$ for any smooth $x_{i}$'s,
$[L\varphi]$ denotes the result
of the action of $L$ on the function $\varphi$, with the convention
$$
[(\alpha\partial^{-1}\beta)\varphi](x)=\alpha(x)\int\limits_{0}^{x}\beta
\varphi;$$
in (6), $\lambda$ is a formal parameter.
\end{proposition}

{\bf Proof.} By induction. (5) is clear for $p=0,1$; assume it is true
for $p$, then
$$
 {\cal L}^{p+1}={\cal L}{\cal L}^{p}=\sum_{i=1}^{N}\sum_{k=0}^{p}[[{\cal L}
[{\cal L}^{k}\varphi_{i}]]\partial^{-1}[{\cal L}^{*(p-k-1)}\psi_{i}] +
\sum_{i=1}^{N}\varphi_{i}\partial^{-1}[{\cal L}^{* p}\psi_{i}] +{\rm
diff.\  operator},
$$
this follows from the identity
$$
{\cal M}{\cal N}=\sum_{j}[{\cal M}c_{j}]\partial^{-1}d_{j} +\sum_{i}
a_{i}\partial^{-1}[{\cal N}^{*}b_{i}]+ {\rm diff.\  operator},
$$
for ${\cal M}={\cal M}_{+} + \sum a_{i}\partial^{-1}b_{i}$ and
${\cal N}={\cal N}_{-} + \sum c_{j}\partial^{-1}d_{j}$;
for a proof of this identity cf. \cite{EOR}.
	We define now the following flows on ${\cal L}$:
$$
 \partial_{i,s}{\cal L}= [\sum_{k=0}^{s-1}
[{\cal L}^{k}\varphi_{i}]\partial^{-1}[{\cal L}^{*(s-1-k)}\psi_{i}],
 {\cal L}],\eqno (7a)
$$
$$
 \partial_{\infty,s}{\cal L}= [({\cal L}^{s/n})_{+},{\cal L}].\eqno (7b)
$$
These flows are defined on the phase space ${\cal P}$ because of the
following identities:
$$
\partial_{i,s}{\cal L}= \sum_{k=0}^{s-1}
\sum_{j=0}^{N}[{\cal L}^{k}\varphi_{i}]
[\partial^{-1}([{\cal L}^{*(s-1-k)}
\psi_{i}]\varphi_{i})]\partial^{-1}\psi_{j} +
\sum_{k=0}^{s-1}\sum_{j=0}^{N}[{\cal L}^{k}\varphi_{i}]
[\partial^{-1}[{\cal L}^{*(s-k}\psi_{i}] -
$$
$$
\sum_{k=0}^{s-1}\sum_{j=0}^{N}\varphi_{i}\partial^{-1}[-\partial^{-1}
([{\cal L}^{k}\varphi_{i}]\psi_{j}
[{\cal L}^{*(s-1-k)}\psi_{i}] - \sum_{k=0}^{s-1}\sum_{j=0}^{N}
[{\cal L}^{k}\varphi_{i}]
\partial^{-1}([{\cal L}^{*(s-1-k)}\psi_{i}]+{\rm diff. operator},
$$
$$
\partial_{\infty,s}{\cal L}= \sum_{i=0}^{N}[({\cal
L})_{+}^{s/n}\varphi_{i}]\partial^{-1}\psi_{i} -
\sum_{i=0}^{N}\varphi_{i}\partial^{-1}[({\cal L}_{+}^{*s/n}\psi_{i}]
+ {\rm diff.\  operator};
$$
so we define the flows $\partial_{i,s}$ on ${\cal P}$ by
$$
\partial_{i,s}{\varphi_{j}}= \sum_{k=0}^{s-1}[[{\cal L}^{k}\varphi_{i}]
\partial^{-1}([{\cal L}^{*(s-1-k)}\psi_{i}]\varphi_{j})] - [{\cal
L}^{s}\varphi_{i}]\delta_{ij}
$$
$$
\partial_{i,s}{\psi_{j}}= \sum_{k=0}^{s-1}[[{\cal L}^{* k}\psi_{i}]
\partial^{-1}([{\cal L}^{(s-1-k)}\varphi_{i}]\psi_{j})] + [{\cal L}^{*
s}\psi_{i}]\delta_{ij},\eqno (8)
$$
which can be translated into

$$
\partial_{i;\lambda}\varphi_{j} = [{\cal L}_{i}(\lambda)\varphi_{j}] -
[(1-\lambda{\cal L})^{-1}\varphi_{i}]\delta_{ij},
$$
$$
\partial_{i;\lambda}\psi_{j} = - [{\cal L}_{i}(\lambda)^{*}\psi_{j}] +
[(1-\lambda{\cal L}^{*})^{-1}\psi_{j}]\delta_{ij},\eqno (9)
$$
$$
\partial_{i,\lambda}P = [{\cal L}_{i}(\lambda),{\cal L}]_{+} \eqno(9bis),
$$
where $\partial_{i;\lambda} = \sum_{s\ge 0}\lambda^{s}\partial_{i,s}$ , and
$$
{\cal L}_{i}(\lambda) =[(1-\lambda{\cal
L})^{-1}\varphi_{i}]\lambda\partial^{-1}
[(1-\lambda{\cal L}^{*})^{-1}\psi_{i}] \eqno (10)
$$
is the generating function for the flows (7), so that $(7a)$ holds.
Similarly, we define the flows $\partial_{\infty,s}$ on $\cal P$ by
$$
\partial_{\infty,s}\varphi_{i} = [({\cal L}^{s/n})_{+}{\varphi_{i}}],
\partial_{\infty,s}\psi_{i} = - [({\cal L}^{*
s/n})_{+}{\psi_{i}}], \quad
\partial_{\infty,s}P = [({\cal L}^{s/n})_{+},{\cal L}]_{+} \eqno (11)
$$
so that $(7b)$ holds. (We could also provide generating formulae
$\partial_{\infty;\lambda}P = [((1-\lambda{\cal
L}^{1/n})^{-1})_{+}, {\cal L}]_{+}$, etc.)

Let us prove now:

\begin{proposition} The flows $\partial_{i,s}$,
$i=1,...,N,\infty$, $s\ge 0$, defined by equations (8), $(9bis)$, (11)
commute pairwise.
\end{proposition}

{\bf Proof.} Let us first assume the zero-curvature conditions
$$
[\pr_{i;\la}-{\cal L}_{i}(\la),\pr_{j;\mu}-{\cal L}_{j}(\mu)]=
[\pr_{i;\la}-{\cal L}_{i}(\la),\pr_{\infty,s}-
({\cal L}^{{s\over n}})_{+}]=0\eqno(12)
$$
are proved. (The relations $[\pr_{\infty,s}-
({\cal L}^{{s/ n}})_{+},\pr_{\infty,t}-
({\cal L}^{{t/n}})_{+}]=0$ are known classically.)
They imply the analogous relations, with $\pr_{i;\la}-{\cal L}_{i}(\la)$
replaced by $\pr_{i;\la}-{\cal L}_{i}(\la)+\delta_{ik}(1-\la{\cal
L})^{-1}$ and $\pr_{j;\mu}-{\cal L}_{j}(\mu)$
replaced by $\pr_{j;\mu}-{\cal L}_{j}(\mu)+\delta_{jk}(1-\mu{\cal
L})^{-1}$, due to $(7b)$ and to $\partial_{i;\la}{\cal
L}=[{\cal L}_{i}(\la),{\cal L}]$, and this shows that the vector fields
commute on the variables $\varphi_{i}$, $\psi_{j}$. Since on the other hand,
they imply that they commute on $\cal L$, they will commute on the
phase space $\cal P$.

Let us turn to the proof of (12).
$$
\pr_{i;\la}{\cal L}_{j}(\mu)=[{\cal L}_{i}(\la)(1-\mu{\cal L})^{-1}
\varphi_{j}-\delta_{ij}(1-\mu{\cal L})^{-1}(1-\la{\cal L})^{-1}\varphi_{j}]
\mu\pr^{-1}[(1-\mu{\cal L}^{*})^{-1}\psi_{j}]
$$
$$
+[(1-\mu{\cal L})^{-1}\varphi_{j}]\mu\pr^{-1}
[-{\cal L}_{i}(\la)^{*}(1-\la{\cal L}^{*})^{-1}
\psi_{j}+\delta_{ij}(1-\mu{\cal L}^{*})^{-1}(1-\la{\cal
L}^{*})^{-1}\psi_{j}].
$$

The terms in $\delta_{ij}$ give $\delta_{ij}[{1\over{\la-\mu}}\big(
-{\la\over{1-\la{\cal L}}}+{\mu\over{1-\mu{\cal L}}}
\big)\varphi_{j}]
\pr^{-1}[{\mu\over{1-\mu{\cal L}^{*}}}\psi_{j}]+
\delta_{ij}[{\mu\over{1-\mu{\cal L}}}\varphi_{j}]\pr^{-1}
[{1\over{\la-\mu}}\break \big(
{\la\over{1-\la{\cal L}^{*}}}-{\mu\over{1-\mu{\cal L}^{*}}}
\big)\psi_{j}]=\delta_{ij}{{\la\mu}\over{\la-\mu}}\{
[{1\over{1-\mu{\cal L}}}\varphi_{j}]\pr^{-1}[{1\over{1-\la{\cal L}^{*}}}
\psi_{j}]-
[{1\over{1-\la{\cal L}}}\varphi_{j}]\pr^{-1}[{1\over{1-\mu{\cal L}^{*}}}
\psi_{j}]\}$ $^{*}$ \footnote{$^{*}$
recall that for a formal series $f(\la,\mu)\in{\bf
C}[[\la,\mu]]$, s.t. $f(\la,\la)=0$, we can define the ratio
${f(\la,\mu)\over{\la-\mu}}\in{\bf C}[[\la,\mu]]$; the expressions in
this computation are understood in this sense.}; this will cancel the
analogous terms appearing in
$\pr_{j;\mu}{\cal L}_{i}(\la)$. So
$$
\pr_{i;\la}{\cal L}_{j}(\mu)-\pr_{j;\mu}{\cal L}_{i}(\la)
=\sum_{r}[{\cal L}_{i}(\la)\alpha_{j}^{(r)}]\pr^{-1}\beta_{i}^{(r)}
-\alpha_{j}^{(r)}\pr^{-1}[{\cal L}_{i}(\la)^{*}\beta_{j}^{(r)}]
$$
$$
-\sum_{s}[{\cal L}_{j}(\mu)\alpha_{i}^{(s)}]\pr^{-1}\beta_{j}^{(s)}
-\alpha_{i}^{(s)}\pr^{-1}[{\cal L}_{j}(\mu)^{*}\beta_{i}^{(s)}],
$$
if ${\cal L}_{i}(\la)$ and ${\cal L}_{j}(\mu)$ are written
${\cal L}_{i}(\la)=\sum_{s}\alpha_{i}^{(s)}\pr^{-1}\beta_{i}^{(s)}$ and
${\cal L}_{j}(\mu)=\sum_{r}\alpha_{j}^{(r)}\pr^{-1}\beta_{j}^{(r)}$, and
this coincides with $[{\cal L}_{i}(\la),{\cal L}_{j}(\mu)]$ (applying
again \cite{EOR}, lemma 2 and the fact that the operators have no
differential part).

$\pr_{i;\la}({\cal L}^{{s\over n}})_{+}
=[{\cal L}_{i}(\la),{\cal L}^{{s\over n}}]_{+}$, so
$-\pr_{i;\la}({\cal L}^{{s\over n}})_{+}
+[{\cal L}_{i}(\la),{\cal L}^{{s\over n}}_{+}]
=[{\cal L}_{i}(\la),{\cal L}^{{s\over n}}_{+}]_{-}$;
$\pr_{\infty,s}({\cal L}_{i}(\la))=[({\cal L}^{{s\over n}})_{+}
(1-\la{\cal L})^{-1}\varphi_{i}]\la\pr^{-1}[(1-\la{\cal L}^{*})^{-1}
\psi_{i}]
-[(1-\la{\cal L}^{*})^{-1}\varphi_{i}]\la\pr^{-1}[({\cal L}^{*{s\over
n}})_{+}(1-\la{\cal L}^{*})^{-1}
\psi_{i}]$, so the second part of (12) follows from \cite{EOR}, lemma 2.
\hfill $\Box$
\bigskip

%
%

\section{Poisson structures.}

We are going to define a Poisson structure on $\cal P$, such that the
natural mapping ${\cal P}\to{\cal L}_{n}$ is Poisson, ${\cal L}_{n}$
being endowed with the second Adler-Gelfand-Dickey structure (AGD2).
The cotangent space to $\cal P$ at $(P,\varphi_{i},\psi_{i})$
is defined as $IOP/IOP_{\le -n}\oplus DOP/\{Q\in DOP |
Q\varphi_{i}=Q^{*}\psi_{i}=0, i=1,...,N\}$, where $DOP=\{\sum_{0\le i\le
i_{0}}a_{i}\pr^{i}, a_{i}\in C^{\infty}({\bf R}), i_{0}<\infty\}
$, and
$IOP=\sum_{i<0}a_{i}\pr^{i}, a_{i}\in C^{\infty}({\bf R})\}$, $IOP_{\le
-n}=\sum_{i\le -n}a_{i}\pr^{i}, a_{i}\in C^{\infty}({\bf R})\}$; the
pairing with the tangent space being given by
$$
\langle(X,Q), (\delta P, \delta\varphi_{i}, \delta\psi_{i})\rangle
={\rm tr}(X+Q)(\delta P+\sum_{i=1}^{N}\delta \varphi_{i}\pr^{-1}\psi_{i}
+\varphi_{i}\pr^{-1}\delta\psi_{i})
$$
$$
={\rm tr} X\delta P+\sum_{i=1}^{n}\int
\delta\varphi_{i}[Q^{*}\psi_{i}]+\delta\psi_{i} [Q\varphi_{i}].
$$

The Hamiltonian vector field on ${\cal L}_{n}$ corresponding to the
covector $(X,Q)$ is $V_{(X,Q)}({\cal L})\break =({\cal L}(X+Q))_{+}{\cal L}-
{\cal L}((X+Q){\cal L})_{+}
$; this formula can be realized defining
$$
V_{(X,Q)}(\varphi_{i})=[(PX+{\cal L}Q)_{+}\varphi_{i}]=[({\cal
L}(X+Q))_{+}\varphi_{i}], \eqno(13a)
$$
$$
V_{(X,Q)}(\psi_{i})=-[((XP)^{*}+(Q{\cal L})^{*})_{+}\psi_{i}]=-[({\cal
L}(X+Q))^{*}_{+}\psi_{i}], \eqno(13b)
$$
$$
V_{(X,Q)}(P)=[({\cal L}(X+Q))_{+}{\cal L}-
{\cal L}((X+Q){\cal L})_{+}]_{+}. \eqno(13c)
$$
The corresponding first structure is given by
$$
V_{(X,Q)}^{(1)}(\varphi_{i})=[(X+Q)_{+}\varphi_{i}],
V_{(X,Q)}^{(1)}(\psi_{i})=-[(X+Q)^{*}_{+}\psi_{i}],
V_{(X,Q)}^{(1)}(P)=[{\cal L},X+Q]_{+}+[Q,{\cal L}]\eqno(14)
$$
Let us show:

\begin{proposition}
Formulae (13) and (14) define compatible Poisson
structures on $\cal P$, such that the map ${\cal P}\to{\cal L}_{n}$ is
Poisson with ${\cal L}_{n}$ endowed with the second (resp. first)
Adler-Gelfand-Dickey structures.
\end{proposition}
{\bf Proof.} Let $\tilde{\cal P}$ be the subset of $\cal P$ consisting of
the $(P,\varphi_{i},\psi_{i})$ such that the Wronskians $W(\varphi_{i})$
and $W(\psi_{i})$ are invertible. We have a sequence of maps
$\tilde{\cal P}\to \bar{\cal L}_{n+N}\times
C^{\infty}({\bf R})^{N}\to\bar{\cal L}_{n+N}\times\bar{\cal L}_{N}\to {\cal
L}_{n}$; the first map is defined by
$(P,\varphi_{i},\psi_{i})\mapsto(A,{{\psi'_{1}}\over{\psi_{1}}},\psi_{1}
({\psi_{2}\over\psi_{1}})',...)$; it is a  bijection between open
subsets of the spaces. The second map is the product of the identity and
the Miura mapping. So the composition of the two first maps is
$(P,\varphi_{i},\psi_{i})\mapsto(A,B)$; the inverse of the first map is
$(A,b_{1},...,b_{N})\mapsto(P,\varphi_{i},\psi_{1}(x)
=e^{\int_{0}^{x}b_{1}}\int_{0}^{x}e^{\int_{0}^{y}b_{2}-b_{1}}dy,...)$;
$B$ is equal to $(\pr+b_{1})...(\pr+b_{N})$. The last map is $(A,B)\mapsto
AB^{-1}$.

If $\bar{\cal L}_{n+N}$ and $\bar{\cal L}_{N}$ are endowed respectively
with AGD2 and its opposite, the last map is Poisson; this follows from
\cite{Di}, add. remark, or from the fact that each $\bar{\cal L}_{p}$ is a
union of symplectic leaves of the group ${\cal L}$ of \cite{EKR3},
and that the map $S\times S'\to G$, $(g,h)\mapsto gh^{-1}$ in any
Poisson-Lie group $G$ is Poisson, $S$ and $S'$ symplectic leaves of $G$
endowed with the induced structure and its opposite. If now $\bar{\cal
L}_{n+N}\times C^{\infty}({\bf R})^{N}$ is endowed with the product of AGD2
and the free field structures, the second map is again Poisson, by the
Kupershmidt-Wilson theorem (\cite{KW}). We wish now to show that the first
map is Poisson, if $\tilde{\cal P}$ is endowed with the structure (13).

Under the first map, the correspondence of tangent spaces is given by
$\delta{\cal L}-\delta A B^{-1}-AB^{-1}\delta B B^{-1}$, and for
cotangent spaces it is $X_{A}=(B^{-1}(X+Q))_{-}$, $X_{B}=-((X+Q){\cal
L})_{-}$. On the other hand, we know from \cite{Ra}, \cite{Kho} that the
action of the Hamiltonian vector
field corresponding to $X_{B}$ on solutions of $B^{*}$ is given by
$V_{X_{B}}(\psi_{i})=-(X_{B}^{*}B^{*})_{+}\psi_{i}$. But
$(BX_{B})_{+}=-((X+Q){\cal L})_{+}+B(B^{-1}(X+Q){\cal L})_{+}$ so since
$B^{*}\psi_{i}=0$, $V_{X_{B}}(\psi_{i})$ coincides with the vector field
of (13).

To obtain the other parts of (13), we write the Hamiltonian vector field
$V_{(X,Q)}$ on $\cal L$; it contains $2N$ terms of the form
$a\pr^{-1}b$, namely the $\varphi_{i}\pr^{-1}V_{(X,Q)}(\psi_{i})$
and the $V_{(X,Q)}(\varphi_{i})\pr^{-1}\psi_{i}$, and we identified the
first ones with what is in (13). It means that
$\sum_{i=1}^{N}(V_{(X,Q)}(\varphi_{i})-$ formula for it in
$(13))\pr^{-1}\psi_{i}$ is a differential operator, and this implies
$(13a)$. $(13b)$ and $(13c)$ are deduced directly. The Poisson and
compatibility of the first structure, given by (14) and the second one
are obtained as usual by consideration of the map
$(P,\varphi_{i},\psi_{i})\mapsto (P+\la,\varphi_{i},\psi_{i})$.
\hfill $\Box$
\bigskip

{\bf Remarks.} 1) It is plausible that the flows $\pr_{i,s}$ preserve
the Poisson structures (13) and (14) on $\cal P$. The most natural
expressions for Hamiltonians governing them would be $\int_{{\bf
R}}[(1-\la{\cal L})^{-1}\varphi_{i}]\psi_{i}$ but they are not defined on
our phase space.

2) Stationary solutions problem. This is the problem of
finding solutions to the stationary flow equation $[{\cal L},
\sum_{i,k}\alpha_{i,k}{\cal L}_{i;k}+\sum_{l}\alpha_{\infty;l}{\cal
L}_{+}^{l}]=0$ (where ${\cal L}_{i}(\la)=\sum_{k}\la^{k}{\cal
L}_{i,k}$). We define a spectral curve for this problem in the
following way. Consider a commuting pair ${\cal
A}=\sum_{i}\alpha_{i}\pr^{-1}\beta_{i}= A^{-1}B$ and ${\cal
B}=P+\sum_{j}\gamma_{j}\pr^{-1}\delta_{j}=C^{-1}D$, $A,...,D,P$
differential operators with leading term
$\pr^{a}$,..., and coefficients in $C^{\infty}({\bf R})$ (we assume
that ${\rm deg }B>{\rm deg }A$). For any
$\la\in{\bf C}$, let $E_{\la}$ be the kernel of $B-\la A$.
 Then ${\rm dim }E_{\la}={\rm deg }B$ for
all $\la$. Consider the map $j:E_{\la}\to{\rm Ker }A$,
$\psi\mapsto {\cal A}\psi-\la\psi$; let $V_{\la}$ be its kernel; so
$V_{\la}\neq 0$. $V_{\la}$ is stable under the action of $\cal B$; let
$\mu_{i}(\la)$ be the list of the eigenvalues of its action, and
$\psi_{\la,\mu_{i}(\la)}$ be a common eigenvector. (Here $\cal A$ and
$\cal B$ act in the way explined in prop. 1.) The system
$$
(B-\la A)\psi_{\la,\mu_{i}(\la)}=0,\quad
(D-\mu_{i}(\la) C)\psi_{\la,\mu_{i}(\la)}=0
$$
is satisfied; we can apply to it the classical method (\cite{BC}) of
successive Euclidean divisions to find an algebraic relation between
$\la$ and the $\mu_{i}(\la)$, and a vector bundle over the curve it defines.
To go further
in the integration of this system, we would need to impose additional
constraints, following the method of \cite{Krconstraints}.
\medskip

{\bf Example.} Consider the case ${\cal L}=\psi_{1}\pr^{-1}\psi_{2}-
\psi_{2}\pr^{-1}\psi_{1}=(\pr^{2}+q)^{-1}$, $\pr^{2}+q$ the operator
with solutions $\psi_{1}$ and $\psi_{2}$ (we assume
$W(\psi_{1},\psi_{2})=1$). We have then a flow $\pr_{t}{\cal
L}=[\psi_{1}\pr^{-1}\psi_{2},{\cal L}]$. Writing $\psi_{1}=e^{\varphi}$,
$\psi_{2}(x)=e^{\varphi(x)}\int_{0}^{x}e^{-2\varphi(x)}$, we find
$\pr_{t}\varphi(x)=(\int_{0}^{x}e^{2\varphi})(\int_{0}^{x}e^{-2\varphi})$.

\section{$N$-wave equations.}

Let $(\beta_{ij}(x))_{1\le i\neq j\le N}$ be a system of smooth functions
of $x=(x_{1},...,x_{N})$ where $x\in{\bf R}^{N}$. Let us
consider the $N$-wave system of equations
$$
{{\pr \beta_{jk}}\over{\pr x_{i}}}=\beta_{ji}\beta_{ik},{\rm\  if\ }
i,j,k  {\rm\ are\ all\  different,\ and\
}(\sum_{\alpha=1}^{N}{\pr\over{\pr x_{\alpha}}})\beta_{ij}=0,{\rm \
for\  all\ } i\neq j\eqno(15)
$$
Let now $\tb_{1i}$, $\tb_{i1}$ ($i=2,...,N$) be
smooth functions on $\bf R$; let $\cal L$ be the Lax operator
$$
{\cal L}=\pr+\sum_{2\le i\le N}\tb_{1i}\pr^{-1}\tb_{i1}.
$$

\begin{proposition} Define $\tb_{ij}(x)=\int_{0}^{x}\tb_{i1}\tb_{1i}$,
for $i,j$ distinct and $>1$. Then the functions $\tb_{\alpha\beta}$ satisfy
equations (15), $\pr/\pr x_{i}$ being replaced by the
the Krichever flow $(7a)$, for $s=1$ and index $i$, and
$\pr/\pr x_{1}$ by $\pr/\pr x$. Moreover, with these definitions of
$\tb_{ij}$ for $i\neq j>1$, the system (15) is equivalent
to the zero-curvature conditions
$$
[\pr_{i}-\tb_{1i}\pr^{-1}\tb_{i1},\pr_{j}-\tb_{1j}\pr^{-1}\tb_{j1}]=0,\quad
{\rm \ for\ } i\neq j.
$$
\end{proposition}

{\bf Proof.} The fact that the $\tb_{\alpha\beta}$ obey the first part
of (15) is clear; for the second part, it follows from the form
of $\cal L$. The last part is obtained directly.
\hfill $\Box$
\bigskip

We now show how equations closely connected to the system (15) can be
obtained from a Chern-Simons-type Lagrangian. Let us set $N=3$ in what
follows. Consider the following functional, depending on
functions $(\beta_{ij})_{i,j=1,2,3}$
$$
S(\beta_{ij})=\int_{{\bf R}^{3}}AdA+{2\over
3}A^{3},
$$
where $A$ is the matrix-valued one-form on ${\bf R}^{3}$ with
coefficients $A_{ij}=\beta_{ij}dx_{i}$. The associated Euler-Lagrange
equations are as usual $dA+[A,A]=0$; this can be written
$$
{{\pr\beta_{jk}}\over{\pr x_{i}}}=\beta_{ji}\beta_{ik} {\rm \ for\ }
i\neq j
$$
Any solution to these equations satisfies $\pr_{l}\pr_{k}{\rm log }\beta_{lk}
=\beta_{kl}\beta_{lk}$ and $\pr_{k}\pr_{l}{\rm log }\beta_{kl}
=\beta_{lk}\beta_{kl}$, hence the function $\phi_{kl}={\rm
log}\beta_{kl}\beta_{lk}$ satisfies the Liouville equation
$\pr_{k}\pr_{l}\phi_{kl}=2e^{\phi_{kl}}$.

Finally, let us make some comments of $\tau$-functions for the system
(15). For any solution to (15), we have a function ${\rm log\ }\tau$
such that $\beta_{ik}\beta_{ki}=\pr_{x_{i}}\pr_{x_{k}}{\rm log\ }\tau$.
This $\tau$-function is connected with the KP one as follows: consider
the KP operator $\pr_{y_{i}}-\pr_{x_{i}}^{2}-u_{i}$, ($y_{i}$ are
additional KP times), with $u_{i}=2\sum_{k=1,...,N, k\neq
i}\beta_{ik}\beta_{ki}$, then $u_{i}=2\sum_{k\neq
i}\pr_{x_{i}}\pr_{x_{k}}
{\rm log\ }\tau = -2\pr_{x_{i}}^{2}{\rm log\ }\tau$, in accordance with the
usual
formulae of the KP theory.

\section{Acknowledgements.}

We are grateful to I. Krichever whose talk at Alushta conference 1994
was the starting point of this work. We would like to thank also
A. Losev and M. Bershadsky for discussions, and M. Pavlov who informed
us about the connection of 3-wave systems with the equations of
hydrodynamic type.

This paper was written during the visit paid
by the first author to the Landau insitute of theoretical physics and
the ITEP (Moscow), to whom he expresses his gratitude for their
invitations. The second and third authors are both supported by ISF grants
MGK000 and  RFFI 94-02-03379 and 94-02-14365. The third author is also
supported by the CNRS.


B.E., V.R.: Centre de Math\'ematiques, URA 169 du CNRS,
Ecole Polytechnique, 91128 Palaiseau,
France

V.R.(permanent address): ITEP, 117259, B. Cheremushkinskaya, Moscou,
Russie

A.O.: P.P. Shirshov Oceanology Institute, ul. Krasikova 23, Moscou
117218, Russie.

\end{document}